\documentclass[epj]{webofc}
\usepackage[varg]{txfonts}   
\usepackage{hyperref}
\usepackage{color}
\usepackage{subcaption}
\usepackage{float}

\newcommand{\matriplex}{\textsc{Matriplex}\xspace}
\newcommand{\cplusplus}{{\small\textsc{C++}}\xspace}
\newcommand{\cuda}{{\small\textsc{CUDA}}\xspace}
\newcommand{\chitwo}{{$\chi^2$}\xspace}
\DeclareRobustCommand{\orderof}{\ensuremath{\mathcal{O}}}

\woctitle{Connecting The Dots / Intelligent Trackers 2017}

\begin{document}
\title{Parallelized Kalman-Filter-Based Reconstruction of Particle Tracks on Many-Core Processors and GPUs}

\author{\firstname{Giuseppe} \lastname{Cerati}\inst{4}\fnsep\thanks{\email{giuseppe.cerati@cern.ch}} \and
            \firstname{Peter} \lastname{Elmer}\inst{2}\fnsep\thanks{\email{peter.elmer@cern.ch}} \and
            \firstname{Slava} \lastname{Krutelyov}\inst{1}\fnsep\thanks{\email{vyacheslav.krutelyov@cern.ch}} \and
            \firstname{Steven} \lastname{Lantz}\inst{3}\fnsep\thanks{\email{steve.lantz@cornell.edu}} \and	
            \firstname{Matthieu} \lastname{Lefebvre}\inst{2}\fnsep\thanks{\email{ml15@princeton.edu}} \and	
            \firstname{Mario} \lastname{Masciovecchio}\inst{1}\fnsep\thanks{\email{mario.masciovecchio@cern.ch}} \and
            \firstname{Kevin} \lastname{McDermott}\inst{3}\fnsep\thanks{\email{kevin.mcdermott@cern.ch}} \and	
            \firstname{Daniel} \lastname{Riley}\inst{3}\fnsep\thanks{\email{daniel.riley@cornell.edu}} \and	
            \firstname{Matev\v{z}} \lastname{Tadel}\inst{1}\fnsep\thanks{\email{matevz.tadel@cern.ch}} \and	
            \firstname{Peter} \lastname{Wittich}\inst{3}\fnsep\thanks{\email{wittich@cornell.edu}} \and	
            \firstname{Frank} \lastname{W\"{u}rthwein}\inst{1}\fnsep\thanks{\email{fkw@ucsd.edu}} \and	
            \firstname{Avi} \lastname{Yagil}\inst{1}\fnsep\thanks{\email{ayagil@physics.ucsd.edu}}	
}

\institute{UC San Diego, La Jolla, CA, USA 92093
\and
           Princeton University, Princeton, NJ, USA 08544
\and
           Cornell University, Ithaca, NY, USA 14853
\and
           Fermilab, Batavia, IL, USA 60510-5011
          }

\abstract{%
For over a decade now, physical and energy constraints have limited clock speed
improvements in commodity microprocessors. Instead, chipmakers have been pushed
into producing lower-power, multi-core processors such as Graphical Processing Units (GPU), 
ARM CPUs, and Intel MICs.
Broad-based efforts from manufacturers and developers have been devoted to
making these processors user-friendly enough to perform general computations.
However, extracting performance from a larger number of cores, as well as
specialized vector or SIMD units, requires special care in algorithm design and
code optimization. 
One of the most computationally challenging problems in high-energy particle
experiments is finding and fitting the charged-particle tracks during event
reconstruction. This is expected to become by far the dominant problem at the
High-Luminosity Large Hadron Collider (HL-LHC), for example. Today the most
common track finding methods are those based on the Kalman filter. Experience
with Kalman techniques on real tracking detector systems has shown that they are
robust and provide high physics performance. This is why they are currently in
use at the LHC, both in the trigger and offline.
Previously we reported on the significant parallel speedups that resulted from
our investigations to adapt Kalman filters to track fitting and track building
on Intel Xeon and Xeon Phi. 
Here, we discuss our progresses toward the
understanding of these processors and the new developments to port the Kalman filter
to NVIDIA GPUs.
}

\maketitle

%
\section{Introduction}
\label{sec:intro}
The Large Hadron Collider (LHC) at CERN is the highest energy collider ever constructed. It accelerates two counter-circulating proton beams and brings them to collision
in four locations around a 27 kilometer ring straddling the border between Switzerland and France. By several measures it is the largest man-made scientific device on the planet. The goal of the LHC is to probe the basic building blocks of matter and their interactions. In 2012, the Higgs boson was discovered by the CMS and ATLAS collaborations~\cite{cmshiggs,atlashiggs}. By measuring the energy and momentum of particles produced by the collision, we can infer the existence of massive particles that were created and measure their properties.  This process is known as event reconstruction and consists of integrating information from different detector components.  Track reconstruction, also known as tracking, is one step in event reconstruction and determines the trajectories of charged particles (``tracks'') from a set of positions of energy deposits within the various layers of our detectors (``hits'').  Tracking, in comparison to other aspects of event reconstruction, is the most computationally time consuming, the most sensitive to increased activity in the detector, and traditionally, the least amenable to parallelized processing. The speed of online reconstruction has a direct impact on how much data can be stored from the 40 MHz collisions rate, while the speed on the offline reconstruction limits how much data can be processed for physics analyses. This research is aimed at vastly speeding up tracking.  

\begin{figure}[h]
\centering
\includegraphics[width=0.5\textwidth]{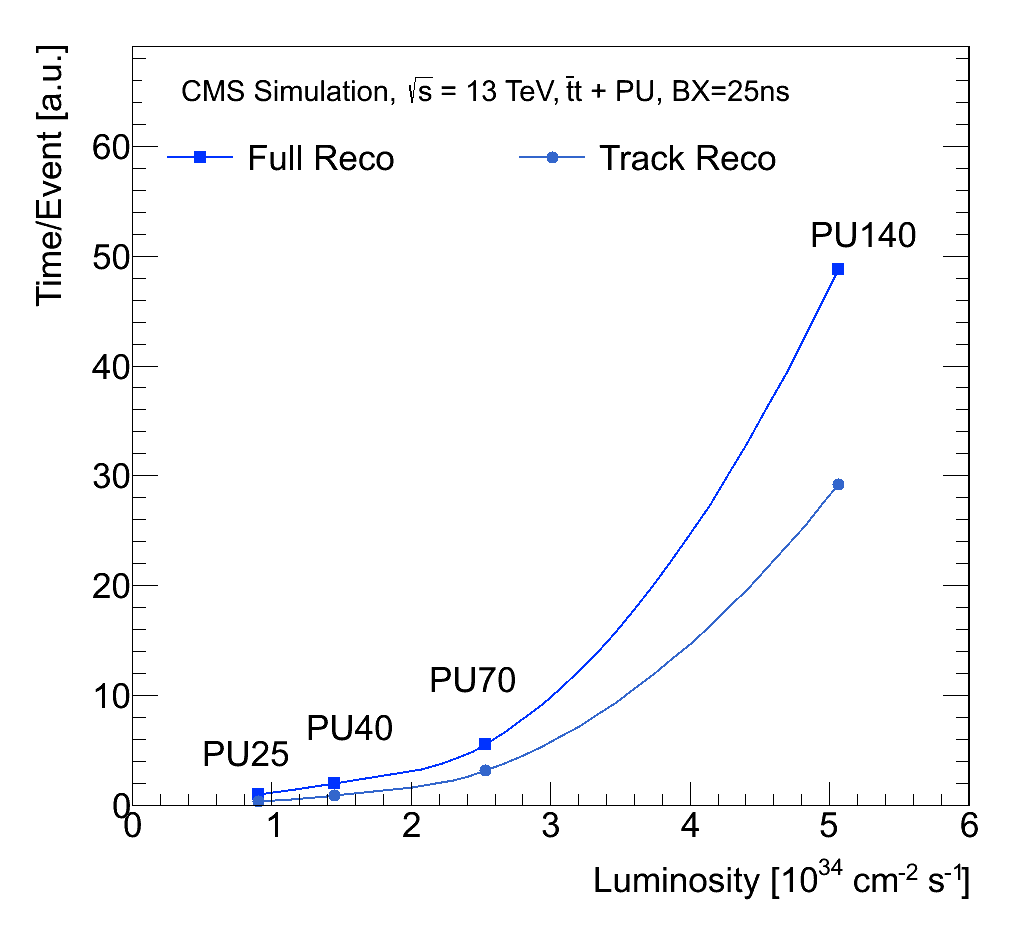}
\caption{CPU time per event versus instantaneous luminosity, for both full reconstruction and the dominant tracking portion. PU25 corresponds to the data taken during 2012, and PU140 corresponds to the HL-LHC era. The time of the reconstruction is dominated by track reconstruction~\cite{vertex}. }
\label{fig:pileup}
\end{figure}

The large amount of time spent in tracking will become even more important in the high-luminosity era of the LHC. The increase in event rate will lead to an increase in detector occupancy (``pile-up'', PU), leading to an exponential gain in the time required to perform track reconstruction, as can be seen in Figure~\ref{fig:pileup}~\cite{vertex}. In the Figure, PU25 corresponds to the data taken during 2012, and PU140 corresponds to the low end of estimates for the HL-LHC era. Clearly our research on tracking performance will become increasingly important during this era.

Accommodating this increasing need for computational power is complicated by the change in computing architectures during the last decade.
Around 2005, the computing processor market reached a turning point: power density limitations in chips ended the long-time trend of ever-increasing clock speeds, and our applications no longer immediately run exponentially faster on subsequent generations of processors. This is true even though the underlying transistor count continues to increase per Moore's law. Increases in processing speed such as those required by the time increases in Figure~\ref{fig:pileup} will no longer come `for free' from faster processors. New, parallel processors instead are aggregates of `small cores' that in total still show large performance gains from generation to generation, but their usage requires reworking of our mostly serial software to exploit these gains. The processors in question include ARM, Graphical Processing Units (GPU) and the Intel Xeon and Xeon Phi.

In what follows, we first update the reader about the algorithms we target and the performance level we now obtain on Xeon and Xeon Phi. We then explain the challenges and the steps we took in porting these algorithms to GPUs along with some initial performance assessments.

\section{Kalman Filter Tracking}
\label{sec:kftrack}
The algorithm we are targeting for parallelized processing is a Kalman Filter (KF) based algorithm~\cite{Fruhwirth:1987fm}. KF-based tracking algorithms are widely used since they are fast and incorporate estimates of multiple scattering in the detector material directly into the trajectory of the particle.  In addition, a combinatorial KF provides an elegant way to resolve ambiguities when considering multiple hits to add to a track~\cite{billoir, Mankel1997}.  Other track finding algorithms, more naturally suited to parallelization and coming from the image processing community, have been explored by others. These include Hough Transforms and Cellular Automata, among others (see, for instance,~\cite{HALYO1}). However, these are not the main algorithms in use at the LHC today, whereas there is extensive understanding on how KF algorithms perform. KF algorithms have proven to be robust and perform well in the difficult experimental environment of the LHC~\cite{vertex}. Rather than abandon the collective knowledge gained from KF algorithms, we wish to extend this robust tool by porting KF tracking to parallel architectures. 

KF tracking proceeds in three main stages: seeding, building, and fitting. Seeding provides the initial estimate of the track parameters based on a few hits in a subset of layers. Realistic seeding is currently under development and will not be reported here. Building then collects additional hits in other detector layers to form a complete track, using the combinatorial KF as a basis for deciding which hits to consider and keep. Track building is by far the most time consuming step of tracking, as it requires branching to explore potentially more than one track candidate per seed after finding compatible hits on a given layer. 
After hits have been assigned to tracks in each layer, a final fit with a KF and smoother can be performed over each track to provide the best estimate of its parameters and to compute a measure of the track quality in terms of a $\chi^2$.

To realize performance gains, we need to exploit two types of parallelism: vectorization and parallelization. Vectorization aims to perform a single instruction on multiple data at the same time by performing the same operation across different data in lockstep.
Parallelization aims to perform different instructions at the same time on different data.
In this work, parallelization is done by distributing the workload between threads, which are concurrent lightweight processes sharing resources (for instance, memory).
The challenge to vectorization is that KF tracking may branch to explore multiple candidates per seed, interfering with the lockstep synchronization required for vectorization performance. The challenge to parallelization is that hit occupancy in a detector is not uniformly distributed on a per event basis, creating the potential for workload imbalances across threads. For these and other reasons, KF tracking cannot be ported in a straightforward way to run in parallel on many-core processors.
Past work by our group has shown progress in porting sub-stages of KF tracking to support parallelism in simplified detectors (see, e.g. our presentations at ACAT2014~\cite{acat2014}, CHEP2015~\cite{chep2015}, and NSS-MIC2015~\cite{nss2015}). As the hit collection is completely determined after track building, track fitting can repeatedly apply the KF algorithm without branching, making this the ideal place to start in porting KF tracking to Xeon and Xeon Phi, with our first results shown at ACAT2014~\cite{acat2014}. 
In more recent works (CHEP2016~\cite{chep2016}, CtD2016~\cite{ctd2016}), we discussed the improvements made to target a more realistic geometry. Profiling and recent computational optimizations were also outlined.

\subsection{Optimized Matrix Library \matriplex}
\label{sec:matriplex}
The computational problem of KF-based tracking consists of a sequence of matrix operations. Unlike~\cite{Fruhwirth:1987fm}, we parameterize the measurement state and the track state in global coordinates with matrices of size $N\times{}N = 3\times{}3$ and $6\times{}6$, respectively.
To allow maximum flexibility for exploring SIMD operations involving many small-dimensional matrices, and to decouple the specific matrix computations from the high level algorithm, we have developed a matrix library, \matriplex, which relies on a matrix-major memory layout (a layout where elements with the same index but from different matrices are stored consecutively).
A more complete description of the \matriplex library can be found in the work we presented at ACAT2014~\cite{acat2014}.
The adaptation of this data structure to GPUs will be discussed in section~\ref{sec:gpu_port}.

\section{Track Building on Xeon and Xeon Phi}
\label{sec:building}
Combinatorial KF track building shares the same core KF calculations as track fitting, but has two major complications for utilizing vectorizable operations compared to track fitting. The first such problem is the ``nHits'' problem: when performing the KF operations, the hit set for a given track candidate in building is undefined and one potentially has to choose between $\orderof{(10^4)}$ hits per layer. The second problem is the ``combinatorial'' problem: when more than one hit on a given layer is compatible with a given input candidate, branching occurs to explore all possible track candidates up to some cutoff number of candidates per seed. 

The key to reducing the nHits problem is to partition the tracks and hits spatially in order to reduce the search space for a given track. We partition space without reference to the detector structure, placing hits in each layer into directionally projective bins. The bins provide a fast look-up of compatible hits in a layer for a given track given the estimate from the KF.


With regard to the combinatorial problem, we first developed track building to add only the best hit out of all compatible hits on a layer for each track candidate. By definition, then, each seed only produces one track and does not require copying of track candidates to explore multiple hits per layer, as in the full combinatorial KF approach.  The best hit is defined as the hit that produced the lowest $\chi{}^2$ increment to the track candidate. The vectorization and
parallelization performance of this ``best hit'' approach were presented at CHEP2015~\cite{chep2015}. After demonstrating the feasibility of track building in the best hit approach, we then allowed for more than one track candidate per seed per layer to be produced, as in the full combinatorial KF approach and presented the results at CtD2016~\cite{ctd2016}.

\subsection{Memory Management}
\label{sec:memmgt}

With the full combinatorial approach in place, we performed extensive studies of the performance of our software, in terms of both the physics performance and the code performance. For the latter, we used the Intel VTune 2016~\cite{vtune} suite of tools to identify bottlenecks and understand the effects of our optimization attempts. In particular, as can be expected, we determined that memory management is of critical importance.


In the full combinatorial approach, if multiple compatible hits are found for a track candidate, it becomes necessary to make copies of that track for each compatible hit. To mitigate the impact from this serial work, we moved copying outside of all vectorizable operations into what we term the ``clone engine''. The clone engine approach only copies the best N track candidates per seed after reading and sorting a bookkeeping list of all compatible hits. This is in contrast to a first attempt at
the combinatorial KF which copied a candidate each time a hit was deemed compatible, then sorting and keeping only the best N candidates per seed after all the possible hits on a given layer for all input candidates were explored.  A more detailed discussion of this work on memory management and impacts on performance was presented at NSS-MIC2015~\cite{nss2015} and at CHEP~2016~\cite{chep2016}

\subsection{Results}
\label{sec:latest}	

\begin{figure}[h]
    \centering
    \begin{subfigure}[t]{0.45\textwidth}
        \includegraphics[width=\textwidth]{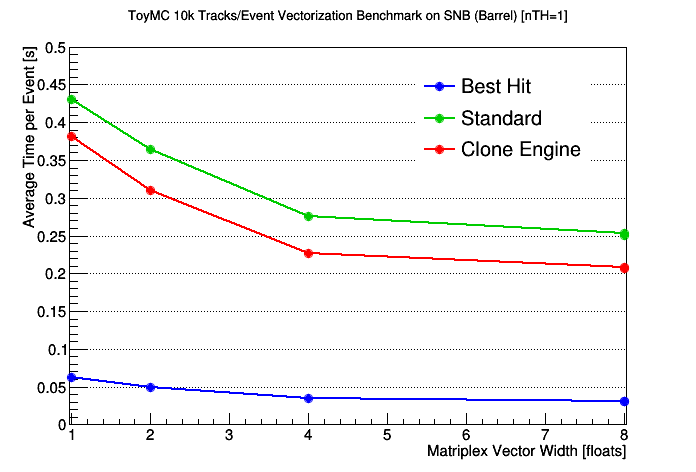}
        \caption{Impact of vector size on performance. Only one thread is enabled.}
        \label{fig:snb_vu}
    \end{subfigure}
    ~ 
    \begin{subfigure}[t]{0.45\textwidth}
        \includegraphics[width=\textwidth]{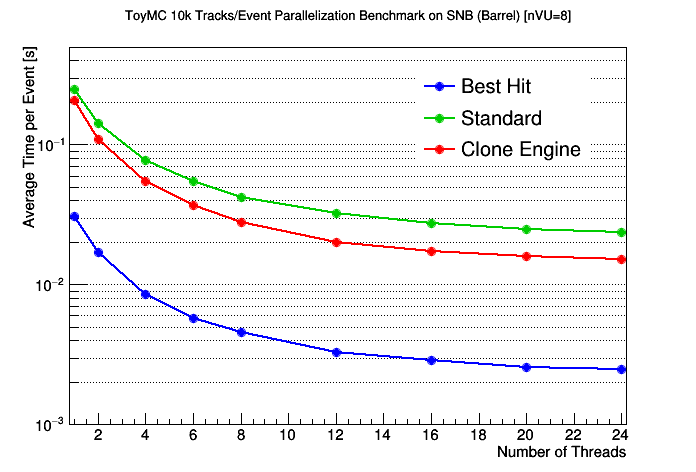}
        \caption{Impact of the number of threads on performance. The full vector size (8 floats) is enabled.}
        \label{fig:snb_th}
    \end{subfigure}
    \caption{Average time per event of $10,000$ tracks on Xeon (Sandy Bridge E5-2620) processors.
    Both plots show results for three different track building approaches. The blue curve is the ``best hit''
    tracking approach. The green and red curves use the full combinatorial approach in
    track building. The red curve moves the copying of tracks outside of the vectorizable
    operations.}
    \label{fig:snb_results}
\end{figure}

We present here the latest vectorization and parallelization benchmarks in track building since CHEP~2016. 
Results are given in average time per event. For each event, 
building proceeds from $10,000$ seeds that are derived from the first three hits of each simulated track in the event. 
We used a simplified detector geometry consisting of equally spaced, fixed length
concentric cylinders.
Tracks are simulated using a set of in-house simulation routines
replicating the detector resolution via Gaussian smearing of the hit positions.
Figure~\ref{fig:snb_results} contains two plots displaying the performance obtained on a computer with two Sandy Bridge multi-core processors (E5-2620 $@$ 2.00GHz).
Plot~\ref{fig:snb_vu} shows the vectorization performance of three different track building approaches, as functions of the number of vector units enabled while using a single thread.
Plot~\ref{fig:snb_th} shows the parallelization performance of the
same track building approaches, as functions of the number of threads enabled
using Intel Thread Building Blocks (TBB)~\cite{tbb_website}.
The Xeon machine we are using has 12 physical cores which appear to be 24
logical cores to the OS due to hyperthreading being enabled. The slight
deviation from ideal scaling before hyperthreading arises from 
load balancing issues between threads.
The large deviation 
from ideal scaling after enabling hyperthreading is due to resource limitations:
the two threads per core are contending for the same instruction pipelines and
data caches. Even so, nearly eightfold speedup is seen for 21 threads.

\begin{figure}[h]
    \centering
    \begin{subfigure}[t]{0.45\textwidth}
        \includegraphics[width=\textwidth]{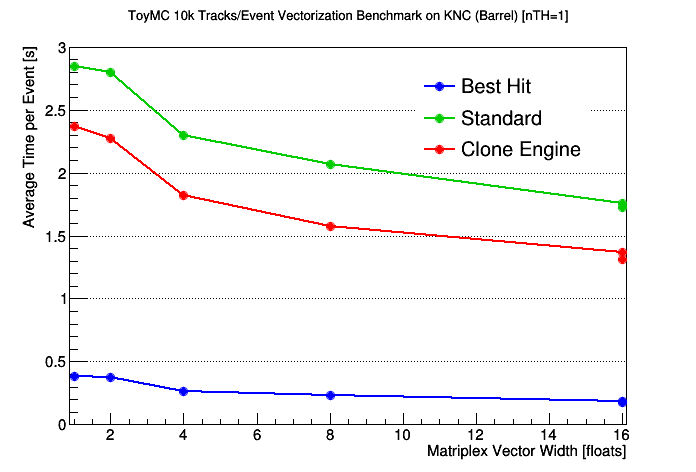}
        \caption{Impact of vector size on performance. Only one thread is enabled.}
        \label{fig:knc_vu}
    \end{subfigure}
    ~ 
    \begin{subfigure}[t]{0.45\textwidth}
        \includegraphics[width=\textwidth]{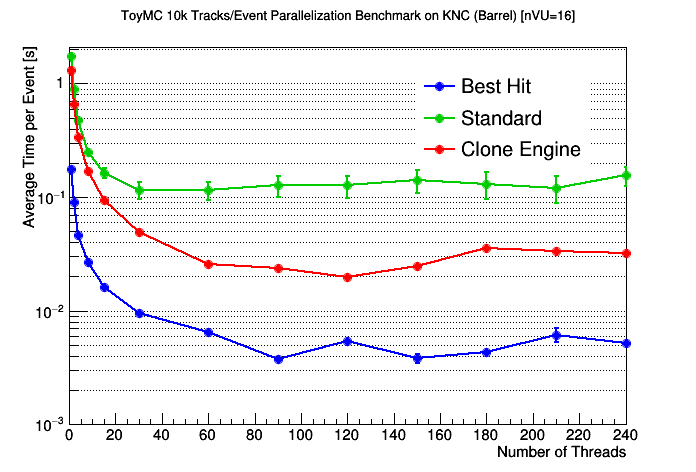}
        \caption{Impact of the number of threads on performance. The full vector size (16
        floating points) is enabled.}
        \label{fig:knc_th}
    \end{subfigure}
    \caption{Average time per event of $10,000$ tracks on Xeon Phi (Knights Corner 7120P) coprocessors.
            Both plots show results for three different track building approaches. The blue curve is the ``best hit'' 
        tracking approach. The green and red curves use the full combinatorial approach in
        track building. The red curve moves the copying of tracks outside of the vectorizable
        operations.}
    \label{fig:knc_results}
\end{figure}

The blue curve is the ``best hit'' approach described previously. Naturally, this approach will have the lowest absolute time in comparison to the two fully combinatorial approaches, the red and green curves. The green curve is the original approach to combinatorial track finding where the copying of the track candidates is inside the vectorizable KF operations. The red curve shows the clone engine approach, which moves the serial work of copying outside of the KF vectorizable operations. As
expected, the clone engine approach has a lower absolute time and higher speedup than the original. It is important to note that while the best hit approach is the fastest, the physics performance has inefficiencies in hit finding and track finding from not being fully able to explore multiple track candidates per seed. Even in our simplified model, this behavior is already apparent, and the best hit approach is expected to become even more inefficient when using more realistic detector effects. For reference, on this simplified problem, the best hit approach approximately has as an efficiency of 93\% and a fake rate of 3\%, while the combinatorial approaches have efficiencies over 99\% and fake rates under 1\%. (Candidates matching fewer than 7 hits of a simulated track count as neither successes nor fakes.)

Figure~\ref{fig:knc_results} shows the same plots as Figure~\ref{fig:snb_results}, now with Xeon Phi (Knights Corner 7120P).
Figure~\ref{fig:knc_vu} shows the vectorization results with Xeon Phi which has AVX-512 vector registers that are twice as wide as those in Xeon. With one thread enabled, Xeon Phi sees the same 1.5$\times{}$ to 2$\times{}$ speedup as Xeon. As seen on Xeon, the best hit approach still has the lowest absolute timing.
Figure~\ref{fig:knc_th} displays the 
parallelization performance, with Xeon Phi, with the full vector width of 16 floats enabled using TBB. The Xeon Phi we are using has 61 physical cores requiring 122 independent instruction streams for full utilization, due to the fact that the Xeon Phi issues instructions for a given thread every other clock cycle. Therefore, to keep a given physical core busy every clock cycle, one has to schedule two threads per core alternating in instruction execution. A form of hyperthreading is also
present on Xeon Phi, yielding a total of 244 hardware threads (logical cores). Overall, a factor of about 30$\times{}$ speedup is observed. 

\section{Porting KF based algorithms to the GPU}
\label{sec:gpu_port}

The challenges to port the Kalman Filter to the GPU are similar in principle to those
on x86 multi-core platforms. The parallelization and vectorization issues we
encounter are only exacerbated by the fine grain parallelism required by
graphical processors.
In particular, the large core count of GPUs and their thread scheduling policies
force a large number of threads to be concurrently scheduled.
It also implies that coping with branching is critical, as threads of a
\textit{warp} (a group of 32 consecutive threads) execute the same
instruction at the same time.

We have followed the same incremental strategy to move onto GPU architectures as the one
we adopted for Xeon and Xeon Phi architectures.
We first started by studying and choosing a data structure adapted to
operations on many small matrices.
We then ported our algorithms, starting from track fitting: its simpler
nature made it easier to understand the requirements of porting KF based
algorithms to GPUs.
Armed with this knowledge, we made a first stab at porting track building algorithms.
The best hit approach ---where only the candidate yielding
the best $\chi^2$ is considered at each layer, for each seed ---is the natural one to develop,
as it introduces a limited number of new problems.
The main issue is to understand how to deal with numerous memory indirections.
Pursuing again the same stepwise improvements as those done for x86 processors,
we next looked at the combinatorial algorithm, where multiple candidates per 
seed are considered at each layer. This introduces the additional problem
of branching.

\subsection{Memory Layout}

\begin{figure*}
\centering
\includegraphics[width=0.5\textwidth]{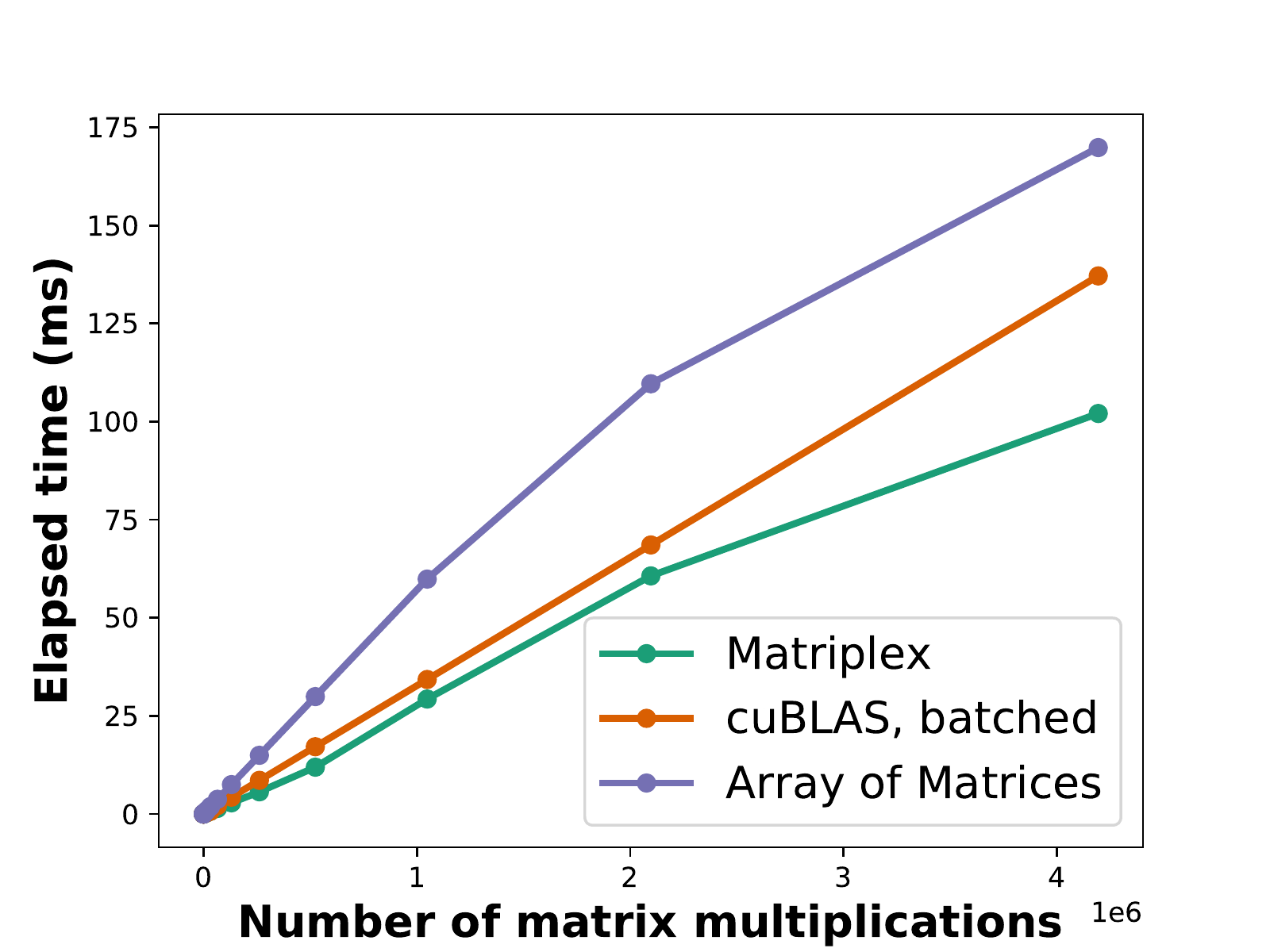}
  \caption{Performance comparison of different memory organizations for 
  multiplying $6 \times 6$ matrices. It shows
  the time as a function of the number of $6 \times 6$ matrix multiplications. The orange
  and purple lines use an array of matrices, while the green line uses the \matriplex
  data structure.}
\label{fig:gplex_perfs}       
\end{figure*}

\matriplex has been found to be the most efficient memory layout for GPU, as it
naturally offers coalescent accesses.
Coalescent accesses occur when, in the case of 4-byte elements, 32 consecutive
threads request 32 consecutive elements from global memory.
In conjunction with aligned memory accesses, it has the desirable effect of
allowing for a single memory transaction instead of 32 separate ones. 
As a global memory (the large off-chip memory) request takes between 200 and 400
clock cycles to complete, this leads to significant gains.
In general, poor performance can be expected unless memory accesses are coalescent.
A key difference with CPU \matriplex is the number of matrices grouped within
such a data structure. Instead of being tailored to the dimension of a hardware
vector unit, it contains as many matrices as possible in order to increase the
parallelism exposed to the device.
Figure~\ref{fig:gplex_perfs} shows some results of a study we conducted to
determine the best memory layout, considering that a large part of the KF
operations involves multiplying small matrices.
In all cases, shared memory is used to reduce the number of global memory
transactions, and this has a dramatic impact on the performance, as much as two orders
of magnitude.
By using the \matriplex data structure, we are able to outperform the finely
tuned and batched cuBLAS approach.
As a results, all our subsequent work has been based on the GPU version of the
\matriplex data structure.

\subsection{Track Fitting on the GPU}

As mentioned, track fitting is suitable for developing a better
understanding of the requirements to obtain performance on an unexplored
hardware architecture, such as GPUs.
Porting from \cplusplus to \cuda is made tedious by the lack of a library
matching the \cplusplus STL and usable from device code.
The existing options (for example NVIDIA Thrust~\cite{nvidia_thrust}) are higher level abstractions to be used from the host and have to be casted to raw pointers when passed to device kernels.
The process of optimizing the code so that it runs fast enough is often orders of magnitude harder than the
process of rewriting to \cuda.
In what follows, we choose the most straightforward way to map computations to
the GPU by using one GPU thread per track during fitting.

\begin{figure*}
\centering
\includegraphics[width=0.5\textwidth]{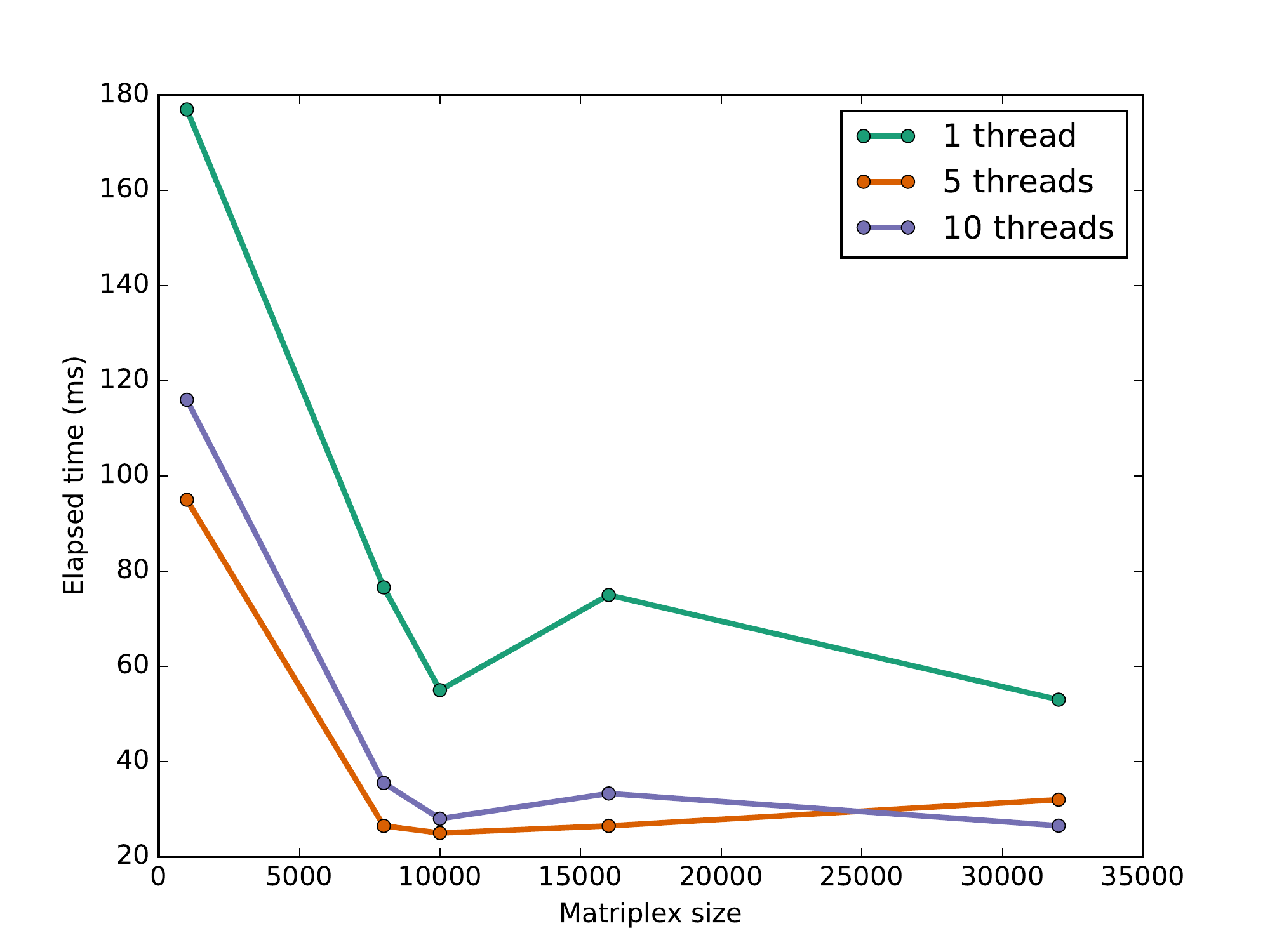}
  \caption{Concurrent fitting processes on the GPU. Different plots show
  different number of CPU threads concurrently streaming events to the GPU. Time
  is plotted as a function of the \matriplex size.}
\label{fig:gpu_concurrent_fitting}       
\end{figure*}

There is a number of places where optimization can occur. The most common
places are the kernels themselves (i.e.\ the \cuda routines) and the data transfers between the host (CPU)
and the device (GPU).
We have observed that optimizing kernels brings significant gains. 
The most beneficial kernel-level optimizations were using read-only
caches, merging kernels to limit the launch overhead of many small kernels, and
favoring registers over shared memory when possible.
These kernel optimizations resulted in a speedup of roughly 8 with respect to a
naive implementation.
Reorganizing data, in particular to fill \matriplex structures, is costly on the GPU because
of the large number of memory indirections it requires.
For track fitting, we have chosen to let this reorganization happen on the CPU,
which is more suited to this task.
By launching kernels asynchronously inside streams, we are able to overlap these
reorganizations with actual computations.
An additional lesson we learned is the need to fill the GPU by fitting tracks
of more than one event concurrently.
Indeed, one event is too small to completely fill the GPU by itself. By using multiple CPU
threads, each of them associated with a list of events to process, significant
improvements are obtained.
Figure~\ref{fig:gpu_concurrent_fitting} shows the impact of concurrent fitting.
For small \matriplex sizes, the kernels perform poorly due to a low compute intensity.
For larger sizes, multiple CPU threads are better able to fill
the GPU, and performance improves by a factor of roughly 3.

\subsection{Track Building on the GPU}

Track building is a more complex problem to deal with on multi- and many-core
architectures. We approach it with two different algorithms. The first one,
``best hit'' considers a single candidate hit for each seed at each layer.
The second one is a combinatorial algorithm, keeping track of multiple
candidates for each seed at each layer, where candidates are ranked by the number of
hits found and by their \chitwo. Note that only the ``clone engine''
approach has been attempted on GPUs: the other combinatorial algorithm involves too many memory
copies to be even considered. 

Compared to track fitting, the ``best hit'' approach increases the complexity by
adding many memory indirections. For each track, hits within $\eta$ bins have to be
considered. This implies frequent reorganizations into \matriplex data structures, making the
overlapping of reorganizations and computations not possible anymore.
Reorganizing hits into \matriplex is consequently done on the GPU\@. As in track
fitting, the parallelization relies on having one GPU thread per candidate.
Modular and nested data structures can fragment data transfers, even if they
are asynchronous, and will in most cases have to be redesigned to a shallower hierarchy.
A data structure with multiple layers also forces the inconvenience of having 
to use additional arrays to keep track of pointers to GPU memory.
As a side note, newer versions of \cuda offer unified memory as a way to
transparently access data that might be located either in the main RAM or in the GPU
global memory. The cost of such a convenience has been greatly reduced over
the past years, but the frequency of these transfers dictates a more controlled
approach, using the knowledge we have to pre-transfer data asynchronously.

\begin{figure}[H]
\centering
\includegraphics[width=0.6\textwidth]{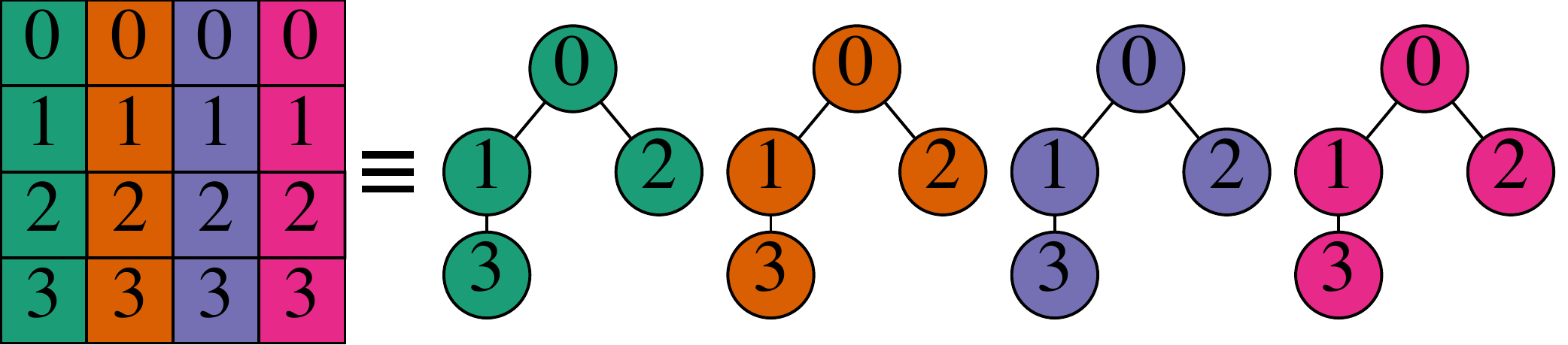}
  \caption{The diagram on the left represents the shared memory layout for storing a seed's candidates. Assuming a maximum of 4 candidates per seed, each column represents a heap to store potential new hits for a track candidate.
  Each different color represents a different initial track candidate of the seed at a given layer. Numbers represent indexes of the heap's elements. Memory is consecutive in the vertical direction. The diagram on the right shows an equivalent representation using binary trees.}
\label{fig:shmem_cands}       
\end{figure}

\begin{figure}[H]
\centering
\includegraphics[width=0.4\textwidth]{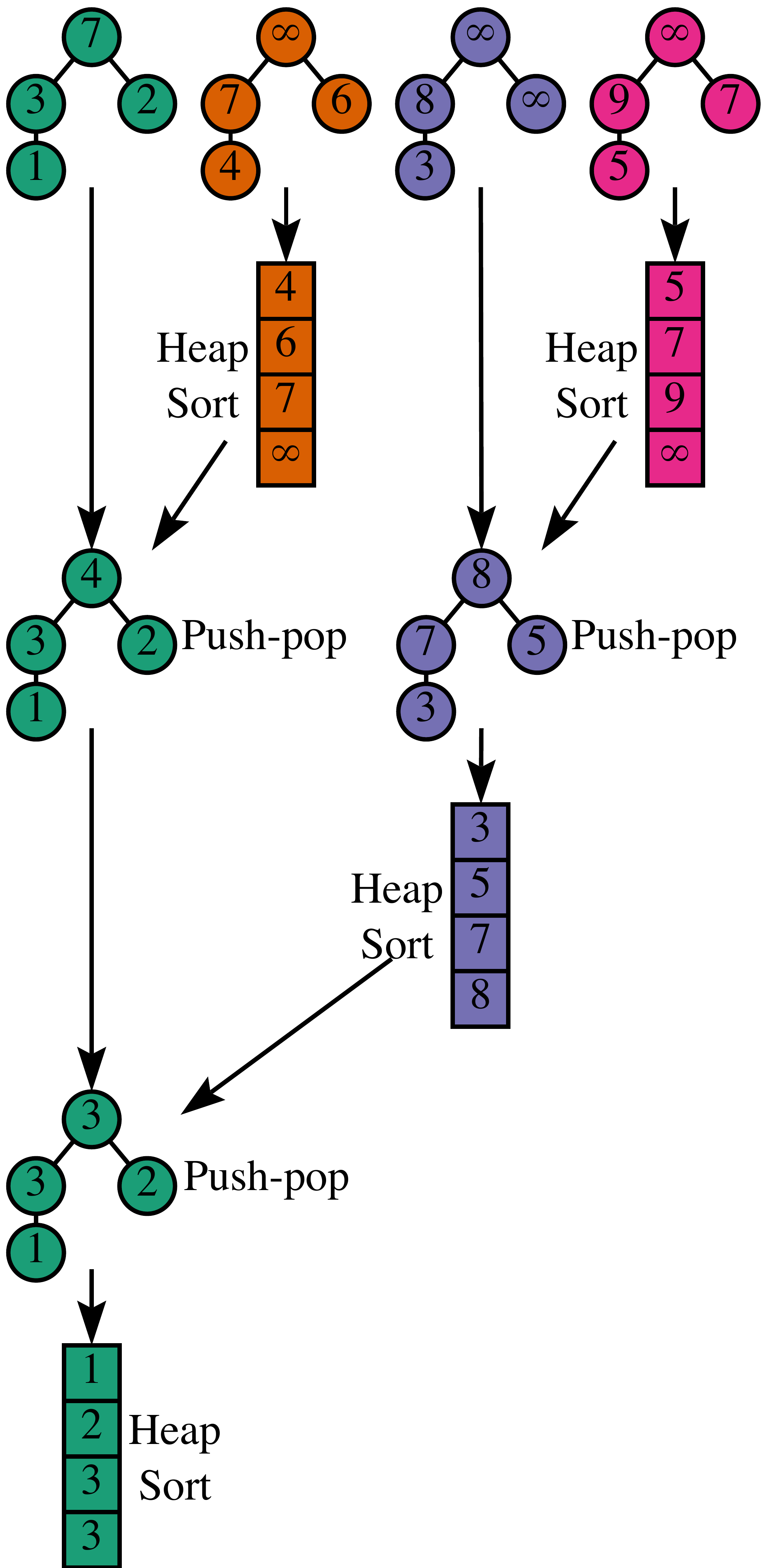}
  \caption{Reducing candidates in shared memory for a single seed. 
  At the end of each layer, the new candidates computed from the starting 4
  candidates are sifted to a new set of 4 candidates. $\infty$ signs represent
  guards for the heap based algorithm. Numbers are $\chi^2$ values
  chosen to be illustrative.
  }
\label{fig:shmem_reduce}       
\end{figure}

The combinatorial approach generates even more complications when aiming at
performant GPU computations. Having multiple candidates per seed at each layer
means that conditional branches are encountered frequently. Since warps always schedule the
same instruction at a given time, branches serialize the warp execution when the
condition differs for each thread of the warp.
We address this issue by defining a data structure to store information
about multiple candidates. 
While on x86 processors standard \cplusplus vectors are used, on GPUs the
underlying choices are more limited and boil down to C-arrays.
Our implementation takes the form of a heap based approach, often used to
implement priority queues. Details are shown in Figure~\ref{fig:shmem_cands}.
A heap is a complete binary tree and can be implemented as a 0-based array where an element at index $i$ has children at indexes $2i+1$ and $2i+2$.
This data structure uses shared memory to avoid extraneous requests to global memory. 
At the beginning of each layer, a heap is assigned to each track candidate to 
store the best suited hits. Each heap is accessed by a single thread, and
elements are consecutive in the vertical direction to avoid bank conflicts. 
At the end of each layer, new overall candidates for the best track need to be found by sifting all the heaps.
Figure~\ref{fig:shmem_reduce} explains this sifting process. A number of heaps are
sorted and their best elements are push-popped into non-sorted heaps. The
process is repeated until a single heap remains. The last heap contains
information about the bests of all the candidates and is sorted in order to proceed to the next layer. Initially, heaps are filled with guards, allowing a different number of potential hits to be considerate per initial track candidate. The number of potential hits is limited by the statically defined heap size.

\begin{figure}[H]
\centering
    \begin{subfigure}[t]{0.45\textwidth}
      \includegraphics[width=\textwidth]{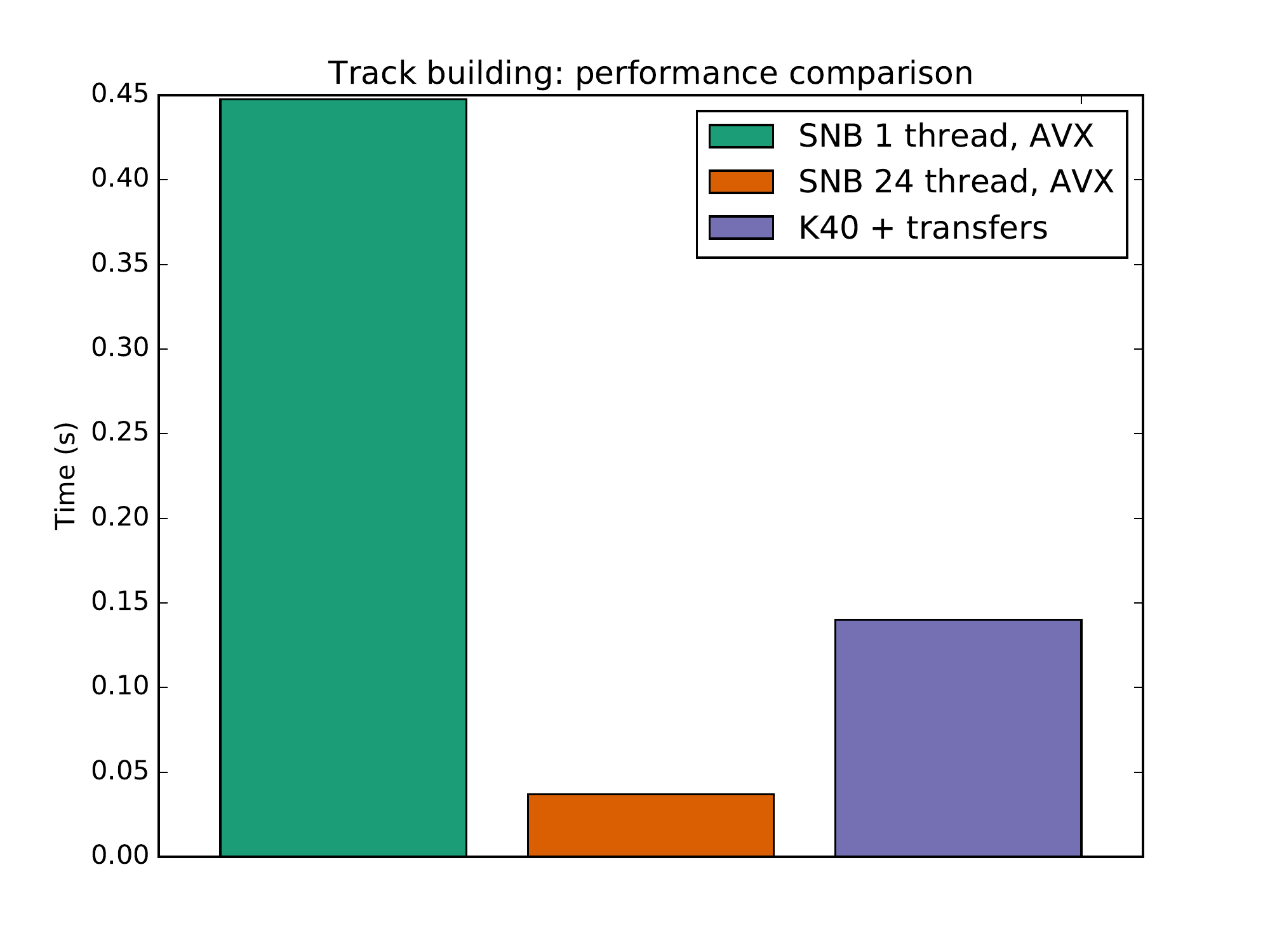}
      \caption{Performance comparison for the ``best hit'' approach.}
      \label{fig:gpu_perfs_bh}       
    \end{subfigure}
    ~
    \begin{subfigure}[t]{0.45\textwidth}
      \includegraphics[width=\textwidth]{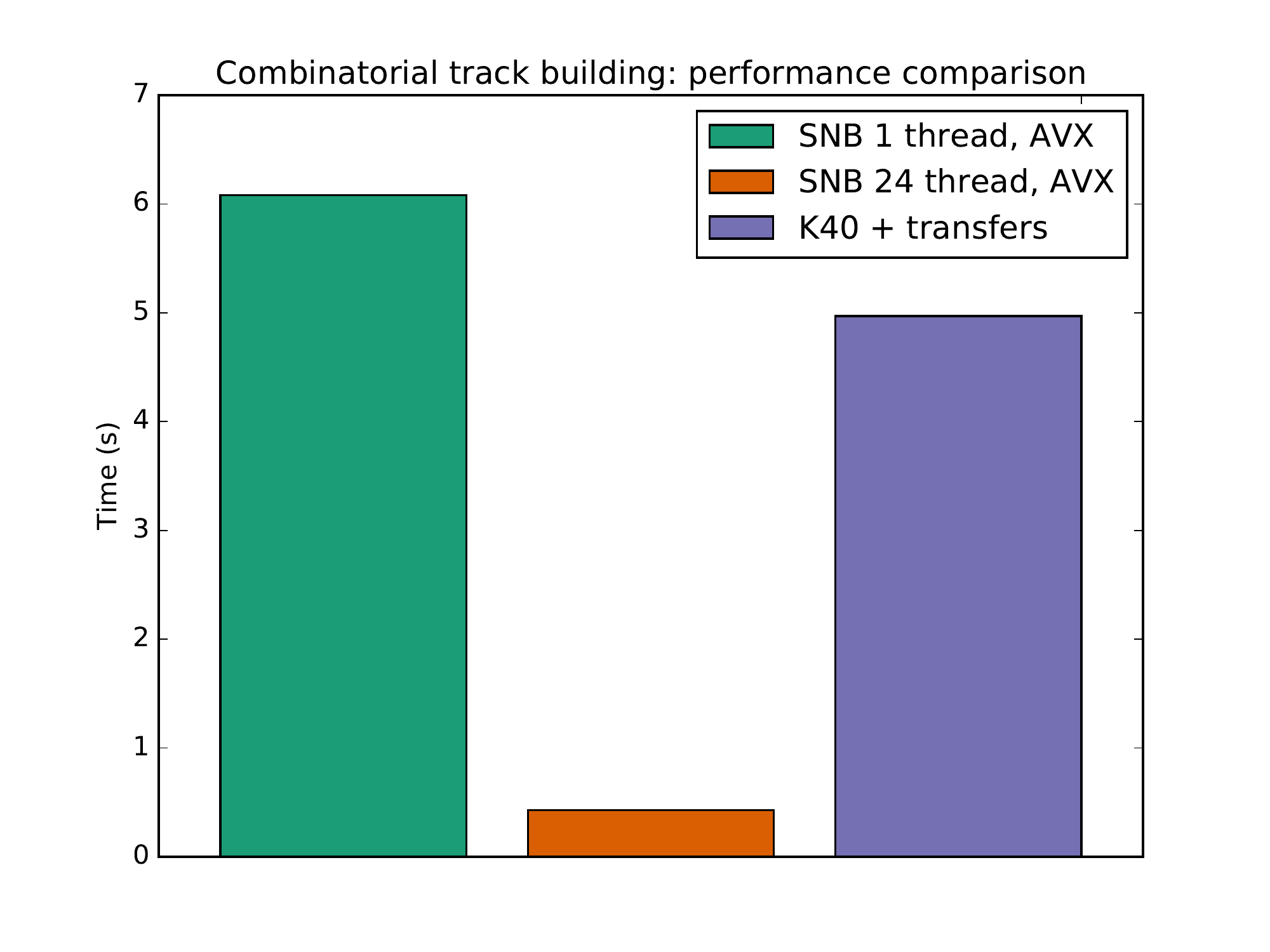}
      \caption{Performance comparison for the combinatoric ``clone engine''
      approach.}
      \label{fig:gpu_perfs_ce}       
    \end{subfigure}
    \caption{Track building performance comparison between the GPU version and
    the optimized Xeon version. The GPU is a NVIDIA K40 and the Xeon a Sandy Bridge E5-2620}
    \label{fig:gpu_perfs_build}
\end{figure}

A large part of the development effort to port track building to GPUs has been to bring the
performance to a level comparable to the one observed on the extensively tuned
Xeon version.
Results for the ``best hit'' approach to track building are shown on
Figure~\ref{fig:gpu_perfs_bh}. Considering the computational time as well as the
data transfer, the GPU version is about 3 times faster than a single-threaded
execution on a Sandy Bridge (SNB) processor but about 4 times slower than an
execution using 24 hyperthreads.
The performance comparison is even more severe for the combinatorial version, as
shown in Figure~\ref{fig:gpu_perfs_ce}. It is a consequence of the large number
of synchronizations and branch predictions.

Several future developments can improve the performance of the GPU code.
First, concurrent events should be adapted to track building because this will help to
fill the GPU's stream multiprocessors more efficiently, as we have seen for track
fitting.
Next, alternative parallelization patterns need to be studied to completely understand what
are the most important performance-driving factors.
For instance, using one GPU thread per seed may reduce the amount
of synchronization required between threads, but will make concurrent event processing
even more critical.
Finally, data transfers need to be drastically improved as they now account for
almost half of the track building wall time.




\section{Conclusion and Outlook}
\label{sec:conclusion}

We have made significant progress in parallelized and vectorized Kalman
Filter-based tracking on multi- and many-core processors.
On Xeon and Xeon Phi, improvements have been made to include a more realistic
geometry. These changes were not discussed here as they are not reflected on the
GPU side, yet. 
On all platforms many challenges remain to fully exploit the computational power
of these highly parallel architectures.
For instance, on the GPU, the bottlenecks we will have to address next are 
data transfers between the host and the device, as well as branching and
synchronization within kernels.
The project has produced promising results; however, much work remains.

\section{Acknowledgment}
\label{sec:ack}
This work is supported by the U.S. National Science Foundation, under the grants PHY-1520969, PHY-1521042, PHY-1520942 and PHY-1120138.

\bibliography{ctd2017}{}

\end{document}